\documentclass[12pt]{article}
\usepackage{amsmath}
\usepackage{amssymb}
\usepackage{epsfig}
\usepackage{graphicx}
\begin{document}

\title{  Dynamics of entanglement in the\\ transverse Ising model   }

\author{Zhe Chang\footnote{e-mail: {\tt changz@mail.ihep.ac.cn}} and Ning Wu\footnote{e-mail: {\tt wun@mail.ihep.ac.cn}} \\
 \\
 {\it Institute of High Energy Physics}\\
 {\it Chinese Academy of Sciences}\\
 {\it P. O. Box 918(4), 100049 Beijing, China}\\}
\date{ }
\maketitle
\begin{abstract}
We study the evolution of nearest-neighbor entanglement in the one
dimensional Ising model with an external transverse field. The
system is initialized as the so called ``thermal ground state" of
the pure Ising model. We analyze properties of generation of
entanglement for different regions of external transverse fields.
We find that the derivation of the time at which the entanglement
reaches its first maximum with respect to the reciprocal
transverse field has a minimum at the critical point. This is a
new indicator of quantum phase transition.

\end{abstract}
\medskip

%%%%%%%%%%%%%%%%%%%%%%%%%%%%%%%%%%%%%%%%%%%%%%%%%%%%%%%%%%%%%%%%%

\section{Introduction}
\par Entanglement has been recognized as an important resource for
quantum information and computation \cite{Book}, and has been the
subject of intense research over the past few years \cite{RMP1}.
Recently, there has been extensive analysis of entanglement in
quantum spin models \cite{RMP2}. Various models have been
considered for entanglement generation, and their static
\cite{20021,20022} as well as dynamical properties
\cite{2004,2005,2009} have been investigated. The reason for this
is that, many of these models can be realized in cold atomic gas
in an optical lattice \cite{Zoller, Hansch}.
\par A special one among these models is the anisotropic XY
model. It can be solved exactly by means of the Jordan-Wigner
transformation \cite{XY}. A lot of works have been done on the
dynamics of entanglement in this model. In Ref.[6], the time
evolution of initial Bell states was studied. Ref.[7] investigated
the dynamics of entanglement of a system prepared in a thermal
equilibrium state, and the system starts evolving after the
transverse field was turned off.
\par In this paper, we analyze a spin-1/2 transverse Ising chain
with a particular separable initial state, the so-called ``thermal
ground state" \cite{20021} for the case of zero field. We study
the generation of nearest-neighbor entanglement as a function of
the external transverse field, and find the time at which the
entanglement reaches its first maximum. It turns out that the
derivation of this time with respect to the reciprocal transverse
field has a minimum at the critical point. This is a new indicator
of quantum phase transition.
\section{Static correlations}
\par The anisotropic XY model is described by the Hamiltonian
\begin{eqnarray}
H=-\frac{\lambda}{4}\sum^N_{i=1}[(1+\gamma)\sigma^x_i\sigma^x_{i+1}+(1-\gamma)\sigma^y_i\sigma^y_{i+1}]-\frac{1}{2}\sum^N_{i=1}\sigma^z_i~,
\end{eqnarray}
where $\sigma^{\alpha}_i$ are the Pauli matrix of the spin at site
$i$ and we assume periodic boundary conditions. We have neglected
a common factor and absorbed the external field into the
reciprocal field $\lambda$. The anisotropy parameter $\gamma$
connects the quantum Ising model for $\gamma=1$. In the range
$0<\gamma\leq1$, the model belongs to the Ising universality class
and undergoes a quantum phase transition at the critical point
$\lambda_c=1$. We will consider the case of $\gamma=1$ in this
paper, namely the transverse Ising model
\begin{eqnarray}
H=-\frac{1}{2}\sum^N_{i=1}(\lambda\sigma^x_i\sigma^x_{i+1}+\sigma^z_i)~.
\end{eqnarray}
\par The Hamiltonian defined in Eq.(2) can be mapped into a
one-dimensional spinless fermion system with creation and
annihilation operators $c^\dag_i$ and $c_i$ via the Jordan-Wigner
transformation
$\frac{\sigma^x_i-i\sigma^y_i}{2}=\prod_{j<i}(1-2c^\dag_jc_j)c_i,~\frac{\sigma^x_i+i\sigma^y_i}{2}=c^\dag_i\prod_{j<i}(1-2c^\dag_jc_j),~\frac{\sigma^z_i}{2}=c^\dag_ic_i-\frac{1}{2}$.
By making use of the transformation
\begin{eqnarray}
\eta_k=\frac{1}{\sqrt{N}}\sum_le^{ikl}(\alpha_kc_l+i\beta_kc^\dag_l)
\end{eqnarray}
further, with $\alpha_k=\frac{\Lambda_k-(1+\lambda \cos
k)}{\sqrt{2[\Lambda^2_k-(1+\lambda \cos
k)\Lambda_k]}},~\beta_k=\frac{\lambda \sin
k}{\sqrt{2[\Lambda^2_k-(1+\lambda\cos k)\Lambda_k]}}$, we get the
fermionic Hamiltonian finally
\begin{eqnarray}
H=\sum_k\Lambda_k\left(\eta^\dag_k\eta_k-\frac{1}{2}\right)~,
\end{eqnarray}
where the spectrum is $\Lambda_k=\sqrt{1+\lambda^2+2\lambda \cos
k}$.
\par The evolution of the system is governed by Eq.(2). In the
Heisenberg picture, the time evolution of the spinless fermion
operator $c_i(t)$ is \cite{2004}
\begin{eqnarray}
c_i(t)=\frac{1}{N}\sum_{k,l}[\cos k(l-i)A(k,t)c_l+\sin
k(l-i)B(k,t)c^\dag_l]~,
\end{eqnarray}
where $A(k,t)=e^{i\Lambda_kt}-2i\beta^2_k\sin\Lambda_kt$,
$B(k,t)=-2i\alpha_k\beta_k\sin\Lambda_kt$.
\par The dynamics of the system also depends on the initial state. We
choose the ``thermal ground state"
$\rho_0=\frac{1}{2}(|N^+\rangle\langle N^+|+|N^-\rangle\langle
N^-|)$ of the pure Ising model (no transverse field is applied) as
the initial state. Here
$|N^+\rangle=|\rightarrow\rangle_1...|\rightarrow\rangle_N$, and
$|N^-\rangle=|\leftarrow\rangle_1...|\leftarrow\rangle_N$ are the
two degenerate ground states of the Ising model with all spins
pointing to the positive (negative) $x$ direction. Note that
$|N^+\rangle$ and $|N^-\rangle$ are the ground states of the
Hamiltonian (2) with $\lambda\to\infty$. When $\lambda$ is finite,
they are even not eigenstates of the Hamiltonian. Thus, the
evolution from this initial state must be nontrivial for finite
$\lambda$. Physically, this can be viewed as an instantaneous,
i.e., idealized sudden quench \cite{2009} in the reciprocal field
$\lambda_1\to\lambda_2$ with $\lambda_1\to\infty$ and
$\lambda_2=\lambda$.
\par Due to the translational symmetry, we need only to consider
averages of the form $\langle c_1...\rangle_0$. Namely, we choose
site-1 as ``the first site", where $\langle...\rangle_0$ denote
the average over the initial state $\rho_0$. We rewrite
$|N^{\pm}\rangle$ in the fermionic representation,
\begin{eqnarray}
|N^\pm\rangle=\left(\frac{1}{\sqrt{2}}\right)^N(1\pm
c^\dag_1)(1\pm c^\dag_2)...(1\pm
c^\dag_N)|0_1,0_2,...,0_N\rangle~,
\end{eqnarray}
where $|0_1,0_2,...,0_N\rangle$ denotes the vacuum of the
fermions, or the state for all spins down. Here we have to pay
attention to the order of operators in this equation. We write
$(1\pm c^\dag_1)$ before operators on all other sites in order to
be compatible with the correlation $\langle c_1...\rangle_0$. In
other words, if another site $i$ is chosen as ``the first site",
we should write the state as
$|N^\pm\rangle=(\frac{1}{\sqrt{2}})^N(1\pm c^\dag_i)(1\pm
c^\dag_{i+1})...(1\pm
c^\dag_{i-1})|0_i,0_{i+1},...,0_{i-1}\rangle$ to calculate
averages like $\langle c_i...\rangle_0$, due to the periodic
boundary conditions.
\par The only single-site averages are
$\langle c_1\rangle_0$ and $\langle c^\dag_1c_1\rangle_0$ (or
their complex conjugate). It is easy to see that $\langle
N^+|c_1|N^+\rangle=\frac{1}{2}=-\langle N^-|c_1|N^-\rangle$, so
$\langle c_1\rangle_0=0$. As to $\langle c^\dag_1c_1\rangle_0$, we
have $\langle
N^+|c^\dag_1c_1|N^+\rangle=\frac{1}{2}\langle0_1|(1+c_1)c^\dag_1c_1(1+c^\dag_1)|0_1\rangle=\frac{1}{2}$.
Similarly, $\langle N^-|c^\dag_1c_1|N^-\rangle=\frac{1}{2}$. So
that, $\langle c^\dag_1c_1\rangle_0=\frac{1}{2}$. We can also see
this point from $0=\langle\frac{\sigma^z_1}{2}\rangle_0=\langle
c^\dag_1c_1-\frac{1}{2}\rangle_0$.
\par Next we consider two-point correlations of the form $\langle
c_1c_l\rangle_0,~l\geq2$. For $l=2$, we have
 \begin{eqnarray}
\langle
N^+|c_1c_2|N^+\rangle&=&\left(\frac{1}{2}\right)^2\langle0_1,0_2|(1+c_2)(1+c_1)c_1c_2(1+c^\dag_1)(1+c^\dag_2)|0_1,0_2\rangle\nonumber\\
&=&\left(\frac{1}{2}\right)^2\langle0_1|(1+c_1)c_1(1-c^\dag_1)|0_1\rangle\cdot\langle0_2|(1+c_2)c_2(1+c^\dag_2)|0_2\rangle\nonumber\\
&=&-\frac{1}{4}~\nonumber.
\end{eqnarray}
Similarly $\langle N^-|c_1c_2|N^-\rangle=-\frac{1}{4}$. So that
$\langle c_1c_2\rangle_0=-\frac{1}{4}$. We can show that $\langle
c^\dag_1c_2\rangle_0=\frac{1}{4}$ in the same manner. For $l>2$,
when $c_l$ or $c^\dag_l$ crosses $(1\pm c^\dag_2)$ to its right
side, this will make it be $(1\mp c^\dag_2)$. But
$\langle0_2|(1\pm c_2)(1\mp c^\dag_2)|0_2\rangle=0$. So we have
obtained all the two-point correlations
\begin{eqnarray}
&&\langle c^\dag_ic_j
\rangle_0=\frac{1}{4}(\delta_{i+1,j}+\delta_{i-1,j}+2\delta_{i,j}),\nonumber\\
&&\langle c_ic_j
\rangle_0=\frac{1}{4}(-\delta_{i+1,j}+\delta_{i-1,j}),\nonumber\\
&&\langle c^\dag_ic^\dag_j
\rangle_0=\frac{1}{4}(\delta_{i+1,j}-\delta_{i-1,j}).
\end{eqnarray}
\par By the same means one can check that the averages involving
three fermion operators all vanish.
\section{Reduced density matrix and evolution of concurrence}
~~~~Making use of the periodic boundary conditions, we can focus
on the reduced density matrix $\rho^{12}(t)$ at time $t$ of the
first and second spins. In the
$\{|\uparrow\uparrow\rangle,|\uparrow\downarrow\rangle,|\downarrow\uparrow\rangle,|\downarrow\downarrow\rangle\}$
basis, the matrix elements reads,
 \begin{eqnarray}
\rho^{12}_{11}(t)&=&\langle\uparrow\uparrow|\rho^{12}(t)|\uparrow\uparrow\rangle\nonumber\\
&=&\langle\uparrow\uparrow|\rho^{12}(t)S^+_1S^-_1S^+_2S^-_2|\uparrow\uparrow\rangle+
\langle\uparrow\downarrow|\rho^{12}(t)S^+_1S^-_1S^+_2S^-_2|\uparrow\downarrow\rangle\nonumber\\
&&+\langle\downarrow\uparrow|\rho^{12}(t)S^+_1S^-_1S^+_2S^-_2|\downarrow\uparrow\rangle+
\langle\downarrow\downarrow|\rho^{12}(t)S^+_1S^-_1S^+_2S^-_2|\downarrow\downarrow\rangle\nonumber\\
&=&tr_{12}(\rho^{12}(t)S^+_1S^-_1S^+_2S^-_2)\nonumber\\
&=&tr(\rho(t)S^+_1S^-_1S^+_2S^-_2)\nonumber\\
&=& \langle c^\dag_1c_1c^\dag_2c_2\rangle_t\nonumber.
\end{eqnarray}
Other matrix elements can be obtained similarly. Thus, we have
\begin{eqnarray}
\rho^{12}(t)= \left(%
\begin{array}{cccc}
  \langle c^\dag_1 c_1c^\dag_2 c_2\rangle_t  & -\langle c^\dag_1 c_1c_2\rangle_t  & \langle c_2^\dag c_2c_1\rangle_t  & \langle c_2c_1\rangle_t  \\
  \\
  -\langle c^\dag_2 c^\dag_1 c_1\rangle_t  &  \langle c^\dag_1 c_1c_2c^\dag_2\rangle_t  & -\langle c_1c^\dag_2\rangle_t & \langle c_1c_2c^\dag_2\rangle_t  \\
  \\
  \langle c^\dag_1 c_2^\dag c_2\rangle_t  & -\langle c_2c_1^\dag\rangle_t  & \langle c_1c^\dag_1c^\dag_2c_2\rangle_t  & \langle c_1c^\dag_1c_2\rangle_t  \\
  \\
  \langle c^\dag_1c^\dag_2\rangle_t  & \langle c_2c^\dag_2c^\dag_1\rangle_t  & \langle c^\dag_2c_1c^\dag_1\rangle_t  & \langle c_1c^\dag_1c_2c^\dag_2\rangle_t  \\
\end{array}%
\right),
\end{eqnarray}
All the elements can be equally evaluated in the Heisenberg
picture, e.g., $\rho^{12}_{11}(t)=\langle c^\dag_1 c_1c^\dag_2
c_2\rangle_t=\langle c^\dag_1(t) c_1(t)c^\dag_2(t)
c_2(t)\rangle_0$. We have mentioned that all the static averages
involving three fermion operators equal zero. From Eq.(5), the
fermion operators in the Heisenberg picture are merely some linear
superpositions of the same set of operators in the Schrodinger
picture. So all matrix elements involving three operators in
Eq.(8) also vanish. Using Wick's theorem, one gets all
non-vanishing elements of $\rho^{12}(t)$ as following
\begin{eqnarray}
&&\rho^{12}_{11}(t)=\langle
  c^\dag_1(t)c_1(t)\rangle_0\langle c^\dag_2(t)c_2(t)\rangle_0-\langle
  c^\dag_1(t)c^\dag_2(t)\rangle_0\langle c_1(t)c_2(t)\rangle_0-\langle
  c^\dag_1(t)c_2(t)\rangle\langle
  c^\dag_2(t)c_1(t)\rangle_0,\nonumber\\
 \nonumber\\
&&\rho^{12}_{22}(t)=\rho^{12}_{33}(t)=\langle c^\dag_1(t)c_1(t)\rangle_0-\rho^{12}_{11}(t),\nonumber\\
\nonumber\\
&&\rho^{12}_{44}(t)=1-\rho^{12}_{11}(t)-2\rho^{12}_{22}(t),\nonumber\\
\nonumber\\
&&\rho^{12}_{14}(t)=\rho^{12*}_{41}(t)=\langle
c_2(t)c_1(t)\rangle_0,\nonumber\\
\nonumber\\
&&\rho^{12}_{23}(t)=\rho^{12*}_{32}(t)=-\langle
c_1(t)c^\dag_2(t)\rangle_0.
\end{eqnarray}
We see that if we can calculate $\langle
  c^\dag_1(t)c_1(t)\rangle_0,~\langle
  c^\dag_1(t)c_2(t)\rangle_0$ and $\langle
  c_1(t)c_2(t)\rangle_0$, then the reduced density matrix is
  totally determined.
\par Using Eq.(5) and (7), we find, in the thermodynamic limit, that
\begin{eqnarray}
\langle c^\dag_1(t)c_1(t)\rangle&=&\frac{1}{8\pi}\int^\pi_{-\pi}
dk\sin k(A^*(k)-A(k))(B(k)-B^*(k)) +\frac{1}{4\pi}\int^\pi_{-\pi}
dk(|A(k)|^2+|B(k)|^2)\nonumber\\
&&+ \frac{1}{4\pi}\int^\pi_{-\pi}
dk\cos k(|A(k)|^2-|B(k)|^2)\nonumber\\
&=&\frac{1}{\pi}\int^\pi_0dk\frac{\lambda\sin^2k\sin^2\Lambda_kt}{\Lambda^2_k}+\frac{1}{2},\nonumber\\
\nonumber\\
\langle c^\dag_1(t)c_2(t)\rangle&=&\frac{1}{8\pi}\int^\pi_{-\pi}
dk(1+\cos2k)(|A(k)|^2-|B(k)|^2)
+\frac{1}{4\pi}\int^\pi_{-\pi} dk\cos k(|A(k)|^2+|B(k)|^2)\nonumber\\
&&+\frac{1}{8\pi}\int^\pi_{-\pi}
dk\sin2k[A(k)-A^*(k)]B^*(k)\nonumber\\
&=& \frac{1}{\pi}\int^\pi_0 dk\frac{\lambda\cos k\sin^2k\sin^2\Lambda_kt}{\Lambda^2_k}+\frac{1}{4},\nonumber\\
\nonumber\\
\langle c_1(t)c_2(t)\rangle&=&\frac{1}{4\pi}\int^\pi_{-\pi}
dk\sin2kA(k)B(k)
 +\frac{1}{8\pi}\int^\pi_{-\pi} dk
[B(k)^2-A(k)^2](1-\cos2k)\nonumber\\
&=&\frac{1}{2\pi}\int^\pi_0dk\sin^2k[\frac{i\sin2\Lambda_kt}{\Lambda_k}-\cos2\Lambda_kt-2\frac{\lambda\sin^2\Lambda_k
t}{\Lambda_k^2}(\cos k+\lambda)]
\end{eqnarray}
Note that $\langle c^\dag_1(t)c_2(t)\rangle$ is indeed real, so
that we have $\rho^{12}_{23}(t)=\rho^{12}_{32}(t)$.\\
\par For
bipartite entanglement, a commonly used measure for arbitrary
states of two qubits is the so-called concurrence \cite{conc}. The
concurrence is defined as
\begin{eqnarray}
C(t)&=&\rm{max}\{0,2\lambda_{max}(t)-tr\sqrt{\rho^{12}(t)\tilde{\rho}^{12}(t)}\},\\
\nonumber\\
\tilde{\rho}^{12}(t)&=&\sigma_y\otimes\sigma_y\rho^{12*}(t)\sigma_y\otimes\sigma_y,
\end{eqnarray}
where $\lambda_{max}$ is the largest eigenvalue of the matrix
$\sqrt{\rho^{12}(t)\tilde{\rho}^{12}(t)}$.
\par In the present case,
\begin{eqnarray}
\rho^{12}(t)=\left(%
\begin{array}{cccc}
  \rho^{12}_{11}(t)  & 0  & 0 &    \rho^{12}_{14}(t) \\
  \\
0  &  \rho^{12}_{22}(t)  & \rho^{12}_{23}(t)  & 0  \\
\\
  0  & \rho^{12}_{23}(t)  & \rho^{12}_{22}(t)  & 0  \\
  \\
 \rho^{12*}_{14}(t)   & 0  &  0  &  \rho^{12}_{44}(t) \\
\end{array}%
\right),
\end{eqnarray}
the four eigenvalues of $\sqrt{\rho^{12}(t)\tilde{\rho}^{12}(t)}$
are of the form
\begin{eqnarray}
\lambda_{1}(t)&=&||\rho^{12}_{14}(t)|+\sqrt{\rho^{12}_{11}(t)\rho^{12}_{44}(t)}|,\nonumber\\
\nonumber\\
\lambda_{2}(t)&=&||\rho^{12}_{14}(t)|-\sqrt{\rho^{12}_{11}(t)\rho^{12}_{44}(t)}|,\nonumber\\
\nonumber\\
\lambda_{3}(t)&=&|\rho^{12}_{22}(t)+
\rho^{12}_{23}(t)|,\nonumber\\
\nonumber\\
\lambda_{4}(t)&=&|\rho^{12}_{22}(t)- \rho^{12}_{23}(t)|.
\end{eqnarray}
\begin{figure}
\begin{center}
\includegraphics[height=15cm,angle=90]{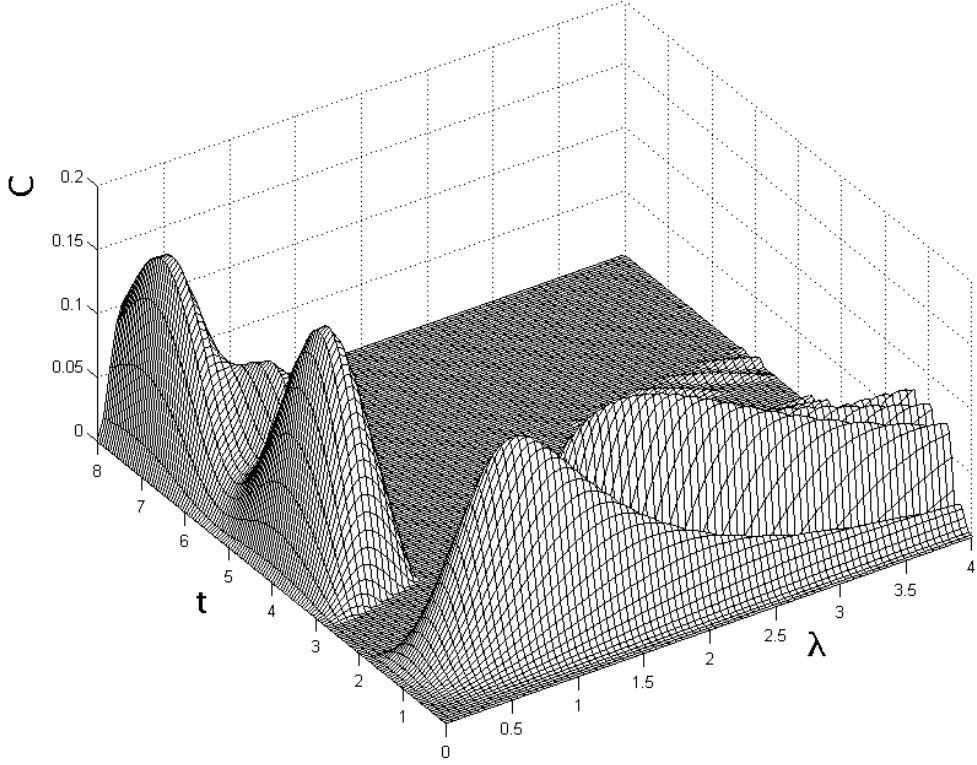}
\hspace{0.5cm}
\end{center}
\caption{\footnotesize{The evolution of bipartite entanglement for
the initial state being the ``thermal ground state" of the Ising
model.}}
\end{figure}
\par In Fig.1, we give a plot of the evolution of entanglement as a function of time $t$ and parameter $\lambda$. At the beginning, the reduced density matrix is
\begin{eqnarray}
\rho^{12}(0)=\frac{1}{4}\left(%
\begin{array}{cccc}
  1  & 0  & 0 &    1 \\
0  &  1  & 1  & 0  \\
  0  & 1  & 1  & 0  \\
1   & 0  &  0  &  1 \\
\end{array}%
\right).
\end{eqnarray}
So that the concurrence equals zero. At any later fixed instant,
the entanglement as a function of $\lambda$ may have different
behaviors in different time intervals. For times before about
$t=2.5$ (Fig.2 (a)), the entanglement maintains its magnitude for
large $\lambda$  with slow oscillations. As $\lambda$ increases,
the height of each peak deceases gradually, and finally vanishes
as $\lambda\to\infty$. This is easy to understand.  For
$\lambda\to\infty$, the Hamiltonian reduces to the pure Ising
model, and the initial state is just the eigenstate of the Ising
model. Thus, only a phase factor contributes to the times evolving
state, which does not change the entanglement. Physically, this
corresponds to $\lambda_1=\lambda_2=\infty$, hence no quench takes
place. For the narrow band $2.47<t<2.69$, there is no entanglement
at all for all $\lambda$. For times greater than around $t=3.0$
(Fig.2 (b)), the entanglement emerges abruptly, but only has
non-vanishing values for very short interval of $\lambda$.
 As the time increases, the
generations of entanglement are different from each other for
different $\lambda$. For $\lambda$ is less than about
$\lambda=0.8$ (Fig.3 (a)), the concurrence will encounter more
than one maximums. In a time region between the first two
maximums, no entanglement appear. Although weak oscillations
emerge after this region, the entanglement can last for a fairly
long time with a not too small magnitude. For $0.80<\lambda<1.07$,
there is only one maximum as the time increases. After a short
time, the entanglement vanishes. For $\lambda>1$ (Fig. 3 (b)),
strong oscillations emerge. The entanglement reaches its maximum
fast, then suddenly decreases to zero. Such a behavior will repeat
again at later times, and finally the entanglement vanishes
completely.
\begin{figure}
\begin{center}
\includegraphics[height=7.93cm,angle=90]{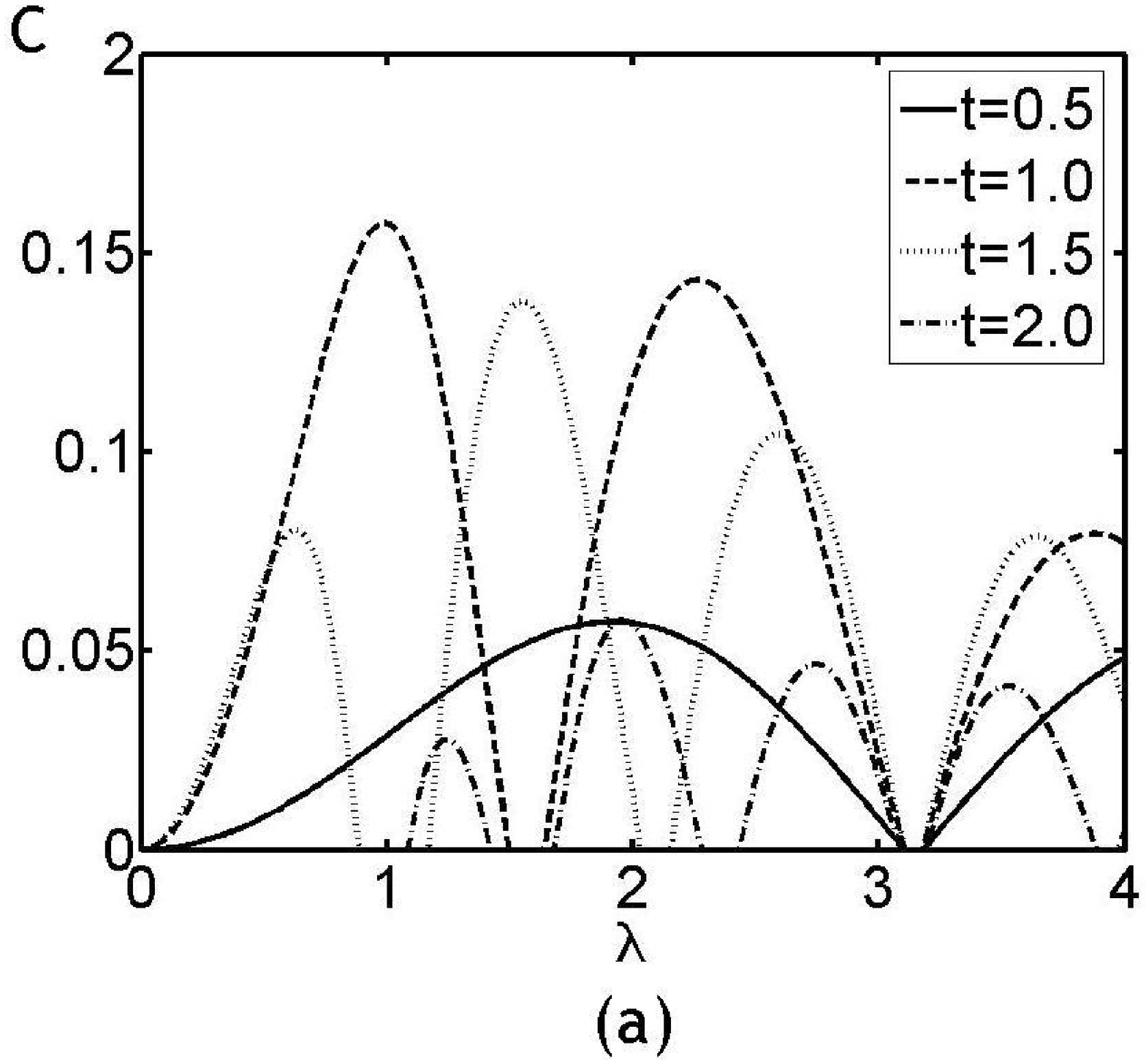}
\includegraphics[height=7.93cm,angle=90]{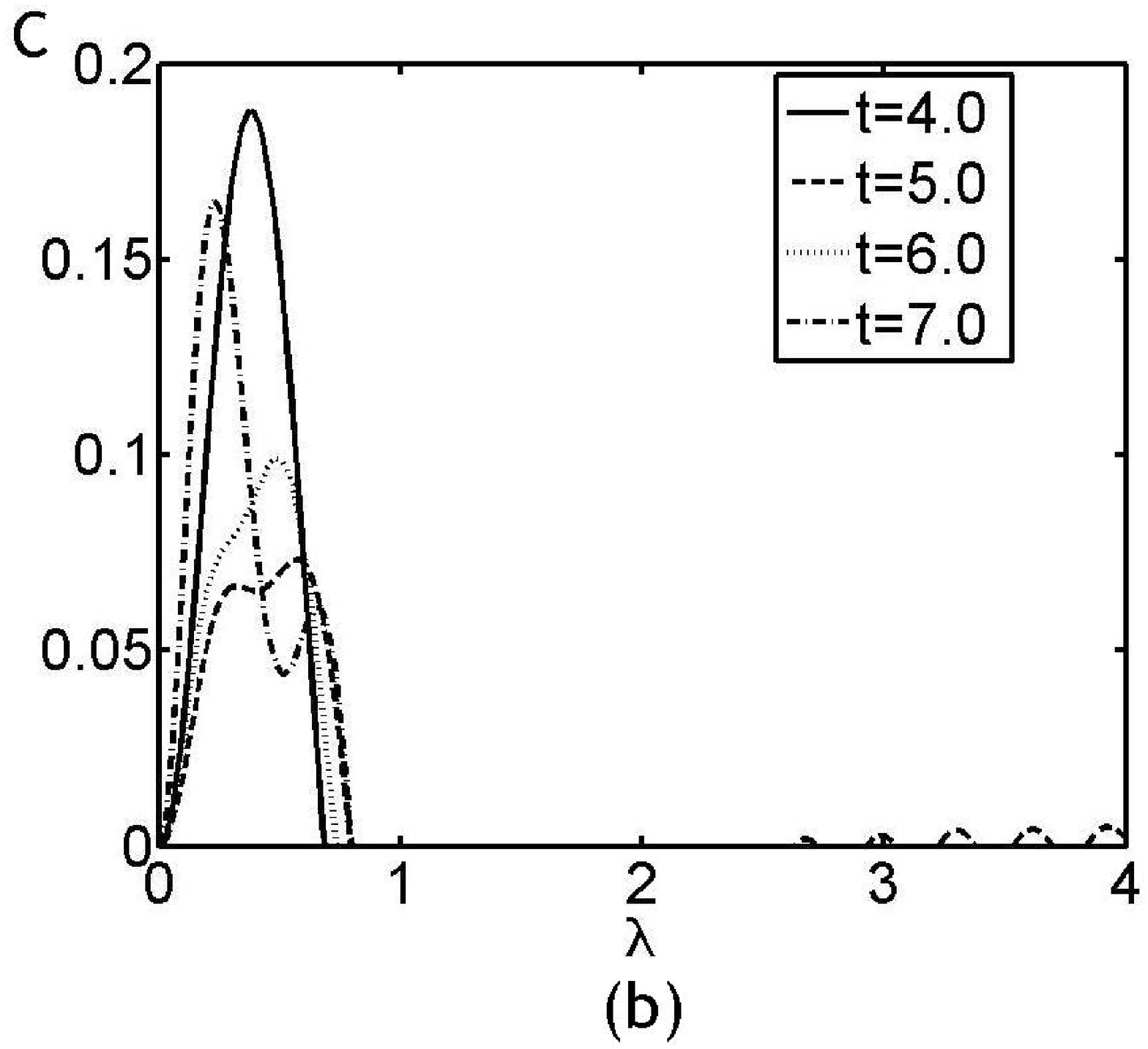} \hspace{0.5cm}
\end{center}
\caption{\footnotesize{(a): The concurrence at different~~~~~~
~~~~~~~~~~~ (b): The concurrence at different times\protect \\
   times $t=0.5,1.0,1.5,2.0$
 as a function of $\lambda$,~~~~~~~~
~~~~~~~~~~~~$t=4.0,5.0,6.0,7.0$ as a function of
$\lambda$,\protect \\ in the region of
$t<2.5$,~~~~~~~~~~~~~~~~~~~~~~~~~~~~~~~~~~~~~~~~~~~~~~~in the
region of  $t>3.0$. }}
\end{figure}
\begin{figure}
\begin{center}
\includegraphics[height=7.93cm,angle=90]{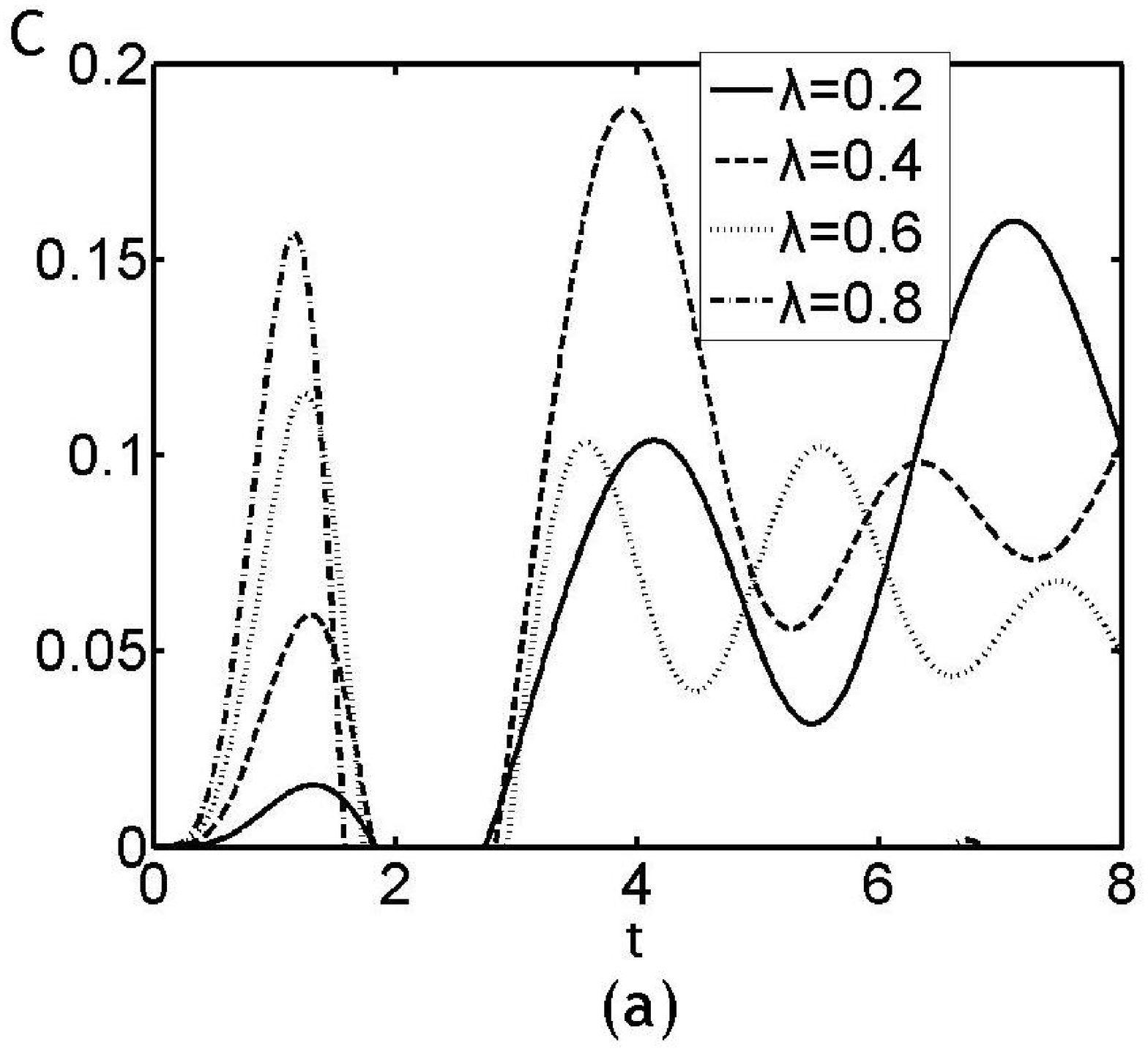}
\includegraphics[height=7.93cm,angle=90]{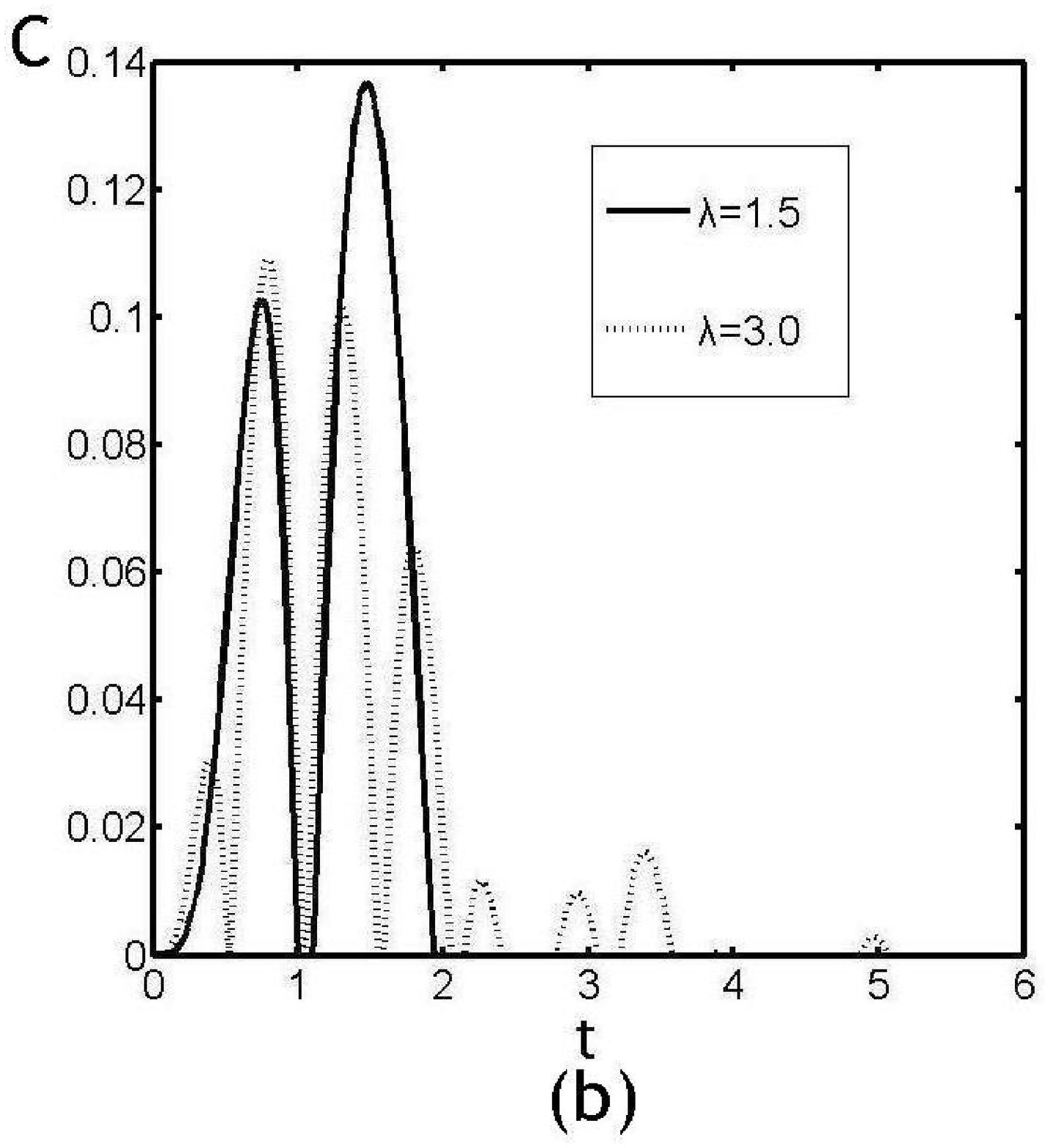}
\hspace{0.5cm}
\end{center}
\caption{\footnotesize{(a): The dynamics of entanglement
~~~~~~~~~~~ (b): The dynamics of entanglement for\protect \\for
different reciprocal fields $\lambda=0.2,0.4,0.6,0.8$,~~~~~~~~~~
 different reciprocal fields $\lambda=1.5,3.0$, \protect \\in
the region of
$\lambda<0.8$,~~~~~~~~~~~~~~~~~~~~~~~~~~~~~~~~~~~~~~~~~~ in the
region of $\lambda>1.07$. }}
\end{figure}
\begin{figure}
\begin{center}
\includegraphics[height=10cm,angle=90]{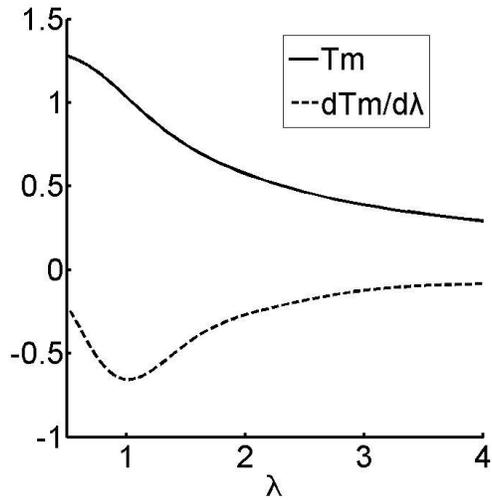}
\hspace{0.5cm}
\end{center}
\caption{\footnotesize{The time at which the entanglement reaches
its first maximum and its derivative.}}
\end{figure}
\par For any $\lambda$, there is a time
$T_{m}(\lambda)$ at which the entanglement reaches its first
maximum (Fig.4). This time is a monotonically decreasing function
of $\lambda$. Taking its derivative with respect to $\lambda$, we
find that the derivative $\frac{dT_m(\lambda)}{d\lambda}$ has a
minimum at the critical point $\lambda=1$. This interesting
property maybe an indicator of the quantum phase transition at the
critical point.
\section{Brief discussions and conclusions}
\par In this paper, we have studied the dynamics of the transverse
Ising chain prepared in the ``thermal ground state" of the pure
Ising model. For different values of the reciprocal field, the
evolution of the entanglement has dramatic distinctions. We also
obtained the times at which the entanglement reaches its first
maximum, which is a monotonically decreasing function of the
parameter $\lambda$. The derivative of this quantity with respect
to the reciprocal field gets its minimum at the critical field.
This is a new indicator of the quantum phase transition. Let us
finally remark that the initial state we choose is simple to be
prepared experimentally. One can first cool the pure Ising system
down to near absolute zero, then turn on some desired transverse
field at $t=0$ to generate the wanted entanglement.
\section*{Acknowledgments}
We would like to thank X. Li for useful discussions. The work was
supported partially by the NSF of China under Grant No. 10875129.


\begin{thebibliography}{99}


\bibitem{Book} M. A. Nielsen and I. L. Chuang, \emph{Quantum Computation
and Quantum Information} (Cambridage University Press, Cambridge,
2000).
\bibitem{RMP1} R. Horodecki, P. Horodecki, M. Horodecki, and K. Horodecki, Rev.
Mod. Phys. \textbf{81}, 865 (2009).
\bibitem{RMP2} L. Amico, R. Fazio, A. Osterloh, and V. Vedral, Rev.
Mod. Phys. \textbf{80}, 517 (2008).
\bibitem{20021} T. J. Osborne and M. A. Nielsen, Phys, Rev. A
\textbf{66}, 032110 (2002).
\bibitem{20022} A. Osterloh, L. Amico, G. Falci, and R. Fazio,
Nature (London) \textbf{416}, 608 (2002).
\bibitem{2004} L. Amico, A. Osterloh, F. Plastina, R. Fazio, and G. M. Palma, Phys, Rev. A
\textbf{69}, 022304 (2004).
\bibitem{2005} A. Sen(De), U. Sen, and M. Lewenstein, Phys, Rev. A
\textbf{72}, 052319 (2005).
\bibitem{2009} H. Wichterich and S. Bose, Phys, Rev. A
\textbf{79}, 060302(R) (2009).
\bibitem{Zoller} D. Jaksch, C. Bruder, J. I. Cirac, C. W. Gardiner, and P. Zoller, Phys, Rev.
Lett. \textbf{81}, 3108 (1998).
\bibitem{Hansch} O. Mandel, M. Greiner, A. Widera, T. Rom, T. W. Hansch, and I. Bloch, Nature (London) \textbf{425}, 937 (2003).
\bibitem{XY} E. Lieb, T. Schultz, and D. Mattis, Ann. Phys. (N.Y.) \textbf{16}, 407 (1961).
\bibitem{conc} W. K. Wootters, Phys, Rev. Lett. \textbf{80}, 2245.
\end{thebibliography}
\end{document}